\documentclass{sigchi-ext}
\usepackage[T1]{fontenc}
\usepackage{textcomp}
\usepackage[scaled=.92]{helvet} 
\usepackage{graphicx} 
\usepackage{balance}  
\usepackage{booktabs} 
\usepackage{ccicons}  
\usepackage{ragged2e} 

\def\plaintitle{Using Conversational Artificial Intelligence to Support Children's Search in the Classroom} 
\def\emptyauthor{}
\def\plainkeywords{Education; Children; Interface; Conversational Search.}

\title{Using Conversational Artificial Intelligence to Support Children's Search in the Classroom}

\numberofauthors{4}

\author{%
  \alignauthor{%
    \textbf{Garrett Allen}\\
    \affaddr{Boise State University} \\
    \affaddr{Boise, ID 83725, USA} \\
    \email{garrettallen@u.boisestate.edu}\\ 
    }
 \alignauthor{%
    \textbf{Jie Yang}\\
    \affaddr{Delft University of Technology} \\
    \affaddr{Delft, The Netherlands} \\
    \email{J.Yang-3@tudelft.nl}\\ 
    } \vfil
 \alignauthor{%
    \textbf{Maria Soledad Pera}\\
    \affaddr{Boise State University} \\
    \affaddr{Boise, ID 83725, USA} \\
    \email{solepera@boisestate.edu}\\ 
    }
 \alignauthor{%
    \textbf{Ujwal Gadiraju}\\
    \affaddr{Delft University of Technology} \\
    \affaddr{Delft, The Netherlands} \\
    \email{u.k.gadiraju@tudelft.nl}\\ 
    }
}

\definecolor{linkColor}{RGB}{6,125,233}
\hypersetup{%
  pdftitle={\plaintitle},
  pdfauthor={\emptyauthor},
  pdfkeywords={\plainkeywords},
  bookmarksnumbered,
  pdfstartview={FitH},
  colorlinks,
  citecolor=black,
  filecolor=black,
  linkcolor=black,
  urlcolor=linkColor,
  breaklinks=true,
}

\begin{document}

\CopyrightYear{2021}
\setcopyright{rightsretained}
\conferenceinfo{CHI'20,}{April  25--30, 2020, Honolulu, HI, USA}
\isbn{978-1-4503-6819-3/20/04}
\doi{https://doi.org/10.1145/3334480.XXXXXXX}
\copyrightinfo{\acmcopyright}

\maketitle

\RaggedRight{} 

\begin{abstract}
We present pathways of investigation regarding conversational user interfaces (CUIs) for children in the classroom. We highlight anticipated challenges to be addressed in order to advance knowledge on CUIs for children. Further, we discuss preliminary ideas on strategies for evaluation.
\end{abstract}



\section{Children and Traditional Web Search}

Seeking information online is a common task for adults and children alike. The information seeking process involves entering a search query into the interface of commercial Web search engines like Google, using a keyboard or other physical input device, exploring search engine result pages (SERP), and selecting the result that satisfies the corresponding need. While these actions are naturally undertaken by adult searchers, when it comes to mainstream search engines, children are known to struggle. This has been revealed by the 
many studies on how children interact with, perceive, and make use of search engines \cite{gossen2016search,gwizdka2017analysis,vanderschantz2017kids}. As an alternative to traditional interfaces, children now turn to voice interfaces to fulfill their search needs \cite{lovato2019hey,garg2020he}.

\section{Children and Conversational Search}

Prior work in HCI has revealed the many benefits of using conversational user interfaces (CUIs) alongside conventional GUIs across different domains~\cite{baier2018conversational,jung2019turtletalk}. CUIs have also been shown to aid human memorability of information consumed in Web search~\cite{qiu2020conversational,qiu2020towards}. With the availability of devices like Alexa or Google Home along with advances in the domain of speech recognition and artificial intelligence, conversational search has emerged as a topic of interest \cite{rosset2020leading,trippas2021spoken}. This is evidenced by the introduction of datasets \cite{ren2021wizard}, evaluation tracks \cite{dalton2020trec}, and research focusing on identifying user needs \cite{zhang2018towards}, as well as frameworks for their evaluation \cite{radlinski2017theoretical}. Much of the current research on conversational search focuses on adults as the main target audience. We argue for the need to look beyond adults and to instead consider other audiences such as children who also frequently turn to this search modality \cite{lovato2015siri}. Children's requirements and expectations for CUIs differ from those of an adult. Still, children remain underserved by existing technology, e.g., children struggle to be understood by voice assistants \cite{landoni2019sonny}.

Investigations into children's interactions and preferences with CUIs remains relatively unexplored, but those into the use of conversational search interfaces at home discovered that children are not well represented in training data and use cases targeted during device design \cite{lovato2019hey}. Interestingly, children did rate the home devices used for the study highly in terms of trustworthiness, friendliness, and safety. When using voice interfaces, children expend extra effort to try and understand the device in addition to asking informational questions \cite{lovato2015siri}. In their study to determine if children are more successful in their searches when using a voice assistant, Landoni et al. \cite{landoni2019sonny} observed that children are easily distracted during interactions, leading to incomplete searches. A realistic conversational approach may encourage a child's attention to remain on the interface.

\section{Children, Conversation, and the Classroom: \\What's Next?}

The obstacles children face when searching are not limited to every day searches occurring in the \textit{home} context; they are also prominent in the \textit{school} context, with the added burden of finding relevant information to support learning. By maintaining a dialogue with the child, CUIs can be used to provide feedback and instruction on how to best perform a search, supporting the search as learning paradigm \cite{collins2016assessing}. Little is currently understood about how conventional Web search systems can be aligned to serve the differing needs of children. As a new modality for search interactions, voice-only conversational search is likely to be subject to similar differences. For instance, children have difficulty effectively navigating SERP in order to find the information they need \cite{duarte2011and,foss2012children}. Rooted in this struggle faced by children with traditional Web search, we then question: How does a CUI for search, i.e., a conversational search interface, designed around voice-only interaction present results in an intelligible and navigable manner?

Current voice interfaces like Alexa respond well to closed-domain questions, e.g., ``what is the weather?", by providing direct, spoken answers. If a query does not have a direct response, how do we design an interface to adequately handle open ended queries? Smartphone devices that make use of Google's voice assistant opt to redirect users to a traditional SERP in such cases. However, this would not be possible in a voice-only interface. How does a voice-only interface allow for users to navigate through results? Can a CUI, by way of clarifying questions \cite{krasakis2020analysing}, ease the burden of selecting results if more than one is presented? Part of the struggle children experience when navigating results is centered on understanding the results themselves and how they can address their information needs. The complexity of a text can affect whether a user is able to comprehend the content presented. Comprehension significantly decreases if a text is too far above a user's reading level \cite{amendum2016push,amendum2018does}. How can we leverage natural language processing methods to determine the reading level of a user in order to provide results that are more likely to be understood? 

Automatic readability assessment (ARA) is a very well researched space \cite{dubay2007smart}. Early approaches often examined shallow textual features, e.g., number of words, syllables per word, and sentence length. With advancements in machine learning, newer solutions to ARA have looked at more complex text-based features, those of lexical and semantic origin \cite{cha2017language,chen2018word}. More recently, with the advent of BERT \cite{devlin2018bert}, ARA solutions have arisen using the contextual information available from BERT's embeddings \cite{imperial2021knowledge}. Therefore, we posit that a search system with a CUI can determine the reading level of a user from their queries, through the use of ARA solutions with BERT. 
BERT could also enable text simplification \cite{sabharwal2021bert,qiang2020lexical}--the process of replacing complex words with those of similar meaning that are easier to understand. To provide a conversational search system that fully addresses a child's search needs in the classroom, we must ensure that a result being read back to them by the CUI uses language they can understand. 

Inspired by the work presented in \cite{beelen2021does,landoni2019sonny}, we surmise that it would be beneficial to 
detect the reading level of children through their spoken queries and to simplify the descriptive text of returned results. In turn, tailoring the vocal presentation of results to the children users can be achieved. Evaluating the automatic detection of a user's reading level in an offline manner requires labelled data, which could be collected via a user study. For the text simplification component, datasets leveraging Wikipedia and Simple Wikipedia 
could be used.

The hurdle of representation in training data is a difficult obstacle to overcome with research involving children as the target audience. Particularly due to the information able to be collected pertaining to children being limited by federal regulations such as COPPA and GDPR \cite{rights2017privacy,federal1998children}. User studies are one way to overcome this barrier, but at the cost of reproducibility as the data collected during such studies cannot be shared. When looking at design considerations for conversational search devices for children, processes like participatory design \cite{robertson2012participatory}, Wizard of Oz \cite{steinfeld2009oz} or cooperative inquiry techniques \cite{druin_cooperative_1999,fails2013methods} that include children as design partners have been considered. 
The latter has been employed in explorations into what a SERP should look like and preferences for traits and personification of voice assistants, according to children \cite{allen2021engage,landoni2020you,yuan2019speech}. A similar process can be undertaken to seek balance between what children want in a conversational search interface and what they need. Would children prefer an interface similar to Google where they simply verbalize their queries and the rest of the system behaves normally? Will they prefer something like a robot companion, personalized to their specific search skills and education level? To clearly contextualize and define answers to these many design questions, further questions around the evaluation of these systems must be considered. The four pillars of search strategy, user group, task, and environment introduced in \cite{landoni2019sonny} as a framework to assess information retrieval systems can serve as a starting point. The inclusion of the fifth pillar -- ``impact of learning" -- will ensure the framework is capable of properly evaluating search as learning using conversational search interfaces.

It is also worth mentioning that taking a one-size-fits-all approach to designing CUIs for children is not necessarily possible, as children possess varying capabilities depending on age and understanding \cite{milton2021infinity}. Furthermore, children are not the only stakeholders when it comes to classroom activities and tools \cite{murgia2019will,murgia2019seven}. Teachers and parents are also involved. What features need to be included as a means to support teachers in their efforts to instruct youth? A dashboard with data on children's search interactions can benefit teachers by providing insights into which students are struggling with each portion of the information seeking process \cite{allen2021casting}. A dashboard of this nature can also serve the role of an evaluation measure for the search system. Through incorporating the teacher's knowledge and expectations for the search tasks, as well as expertise in child development, researchers can ensure that the systems are adequately meeting the needs of all relevant stakeholders.

Using human-in-the-loop design processes and evaluations, conversational search interfaces can be effectively developed to mitigate the known struggles children have with search tools while providing  search as learning scaffolding.

\textbf{Acknowledgments} Work partially funded by NSF Award \# 1763649.


\bibliographystyle{SIGCHI-Reference-Format}
\bibliography{cui_workshop}

\end{document}